\documentclass[a4paper,11pt]{article}
\usepackage{pos}

\usepackage{newunicodechar}

\newunicodechar{ℏ}{\hbar}
\usepackage{textcomp}
\usepackage{amssymb}
\usepackage{tikz}

\title{Unveiling the Quantum Nature of Black Holes: Towards a Proof of Hawking Radiation through Gamma-Ray Observations}

\author*[1]{Atreya Acharyya}
\author[2,3]{Giacomo Cacciapaglia}
\author[1]{Manuel Meyer}
\author[3,4,5,6]{Francesco Sannino}

\affiliation[1]{CP3-Origins, University of Southern Denmark, Campusvej 55, 5230 Odense M, Denmark}
\affiliation[2]{Laboratoire de Physique Th\'eorique et Hautes \'Energies (LPTHE), UMR 7589, Sorbonne Universit\'e \& CNRS, 4 place Jussieu, 75252 Paris Cedex 05, France}
\affiliation[3]{Quantum Theory Center (\ensuremath{\hbar}QTC) at IMADA \& D-IAS, Southern Denmark Univ., Campusvej 55, 5230 Odense M, Denmark}
\affiliation[4]{Dept. of Physics E. Pancini, Universit\`a di Napoli Federico II, via Cintia, 80126 Napoli, Italy}
\affiliation[5]{INFN sezione di Napoli, via Cintia, 80126 Napoli, Italy}
\affiliation[6]{Scuola Superiore Meridionale, Largo S. Marcellino, 10, 80138 Napoli, Italy}

\emailAdd{atreya@cp3.sdu.dk}
\abstract{Hawking’s groundbreaking prediction that black holes emit thermal radiation and ultimately evaporate remains unverified due to the extreme faintness of this radiation for stellar-mass or larger black holes. In this study, we explore a novel observational strategy to search for Hawking radiation from asteroid-mass \emph{black hole morsels}—hypothetical small black holes that may form and be ejected during catastrophic events such as binary black hole mergers. These \emph{black hole morsels} are expected to emit gamma rays in the GeV--TeV range on observable timescales. We analyze data from the \textit{Fermi}-Large Area Telescope coinciding with the well-localized binary black hole merger GW170814, searching for delayed gamma-ray signatures associated with morsel evaporation. While we find no evidence for such emission, we place exclusion limits on morsel masses, ruling out the $4 \times 10^{8}$\,kg scenario at the 95\% confidence level for a total emitted mass of one solar mass. We also outline future directions, including the incorporation of late-time evaporation spikes, systematic application across the growing gravitational wave catalog, and the enhanced discovery potential of next-generation facilities such as the Cherenkov Telescope Array Observatory.}

\ConferenceLogo{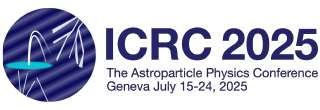}

\FullConference{39th International Cosmic Ray Conference (ICRC2025)\\
 15–24 July 2025\\
Geneva, Switzerland\\}

\begin{document}
\maketitle

\section{Introduction} 
\label{sec:intro}

Black holes are among the most extreme predictions of Einstein’s theory of General Relativity, and their existence has been confirmed through gravitational influences on nearby matter \cite{1998ApJ...509..678G, 2002Natur.419..694S} as well as through the direct imaging of their surrounding environments \cite{2019ApJ...875L...1E}. A striking feature predicted by theory—but still lacking experimental confirmation—is that black holes are not entirely black. In 1974, Stephen Hawking proposed that quantum effects near the event horizon cause black holes to emit thermal radiation. This Hawking radiation is inversely proportional to the black hole’s mass \cite{1974Natur.248...30H}, with the effective temperature scaling as $T \propto 1/M$, so that smaller black holes radiate more intensely at higher energies. For stellar-mass or larger black holes, the radiation is far colder than the cosmic microwave background and thus observationally inaccessible with current instruments. Their evaporation time also scales steeply with mass as $\tau \propto M^3$, so such black holes have lifetimes much longer than the age of the Universe. The strongest observable signal from Hawking radiation is therefore expected only shortly before the final stages of evaporation.

However, if black holes of much smaller mass exist—on the order of $10^8$ kg, roughly the mass of an asteroid—their Hawking radiation would peak in the gamma-ray regime, making them potentially detectable with existing high-energy observatories. Both space-based and ground-based gamma-ray telescopes are capable of detecting photons many orders of magnitude more energetic than visible light. Confirming the existence of Hawking radiation in this energy band would be a profound breakthrough, opening new avenues for exploring quantum gravity.

\begin{figure*}[h]
\centering
\includegraphics[width= \linewidth]{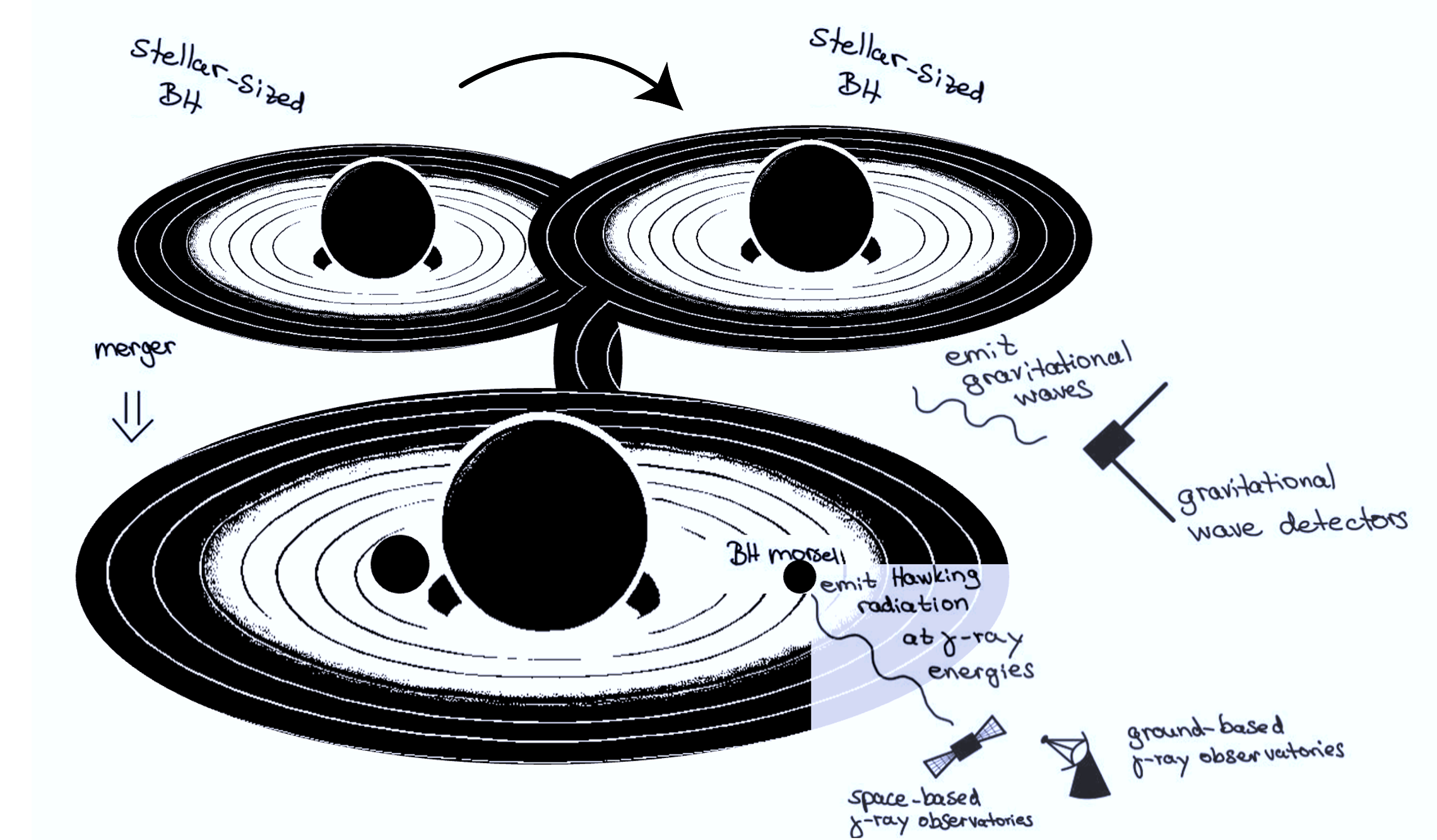}
\caption{The merger of two black holes leading to a gravitational wave event and the creation of a larger black hole, as well as smaller \emph{black hole morsels}. The \emph{black hole morsels} are predicted to emit gamma rays which can be observed with current instruments.}
\label{fig:image1}
\end{figure*}


While such small black holes are not expected to form through standard stellar collapse, following recent theoretical work \cite{2025NuPhB101817021C}  
  we hypothesize that the extreme, non-linear gravitational environment during a merger could produce a multitude of small, evaporating BHs—which we term \emph{black hole morsels}. While speculative, this hypothesis is motivated by various theoretical frameworks beyond GR, including models with extra dimensions, superstring theory, and certain modified gravity theories where BH fragmentation is conceivable (see \cite{Gregory:1993vy, Emparan:2003sy, Chen:2017kpf} and Appendix A of \cite{2025NuPhB101817021C} for a review).
  
These \emph{black hole morsels} are expected to evaporate rapidly via Hawking radiation, emitting gamma-ray photons in a characteristic spectral and temporal pattern. The signal would arrive shortly after the gravitational wave event, with a time delay set by the evaporation timescale of the morsels. If detectable, such a signal would provide a new form of electromagnetic counterpart to gravitational wave events.

In this paper, we search for these signals using data from the \emph{Fermi}--Large Area Telescope (LAT; \cite{LAT}) focusing on the well-localized and relatively nearby binary black hole merger event GW170814. We analyze \emph{Fermi}-LAT data at this location and time, searching for gamma-ray signatures consistent with the predicted \emph{black hole morsel} emission. Furthermore, we statistically rule out certain parameter ranges and impose key constraints on the physics of binary black hole mergers.

\section{Observations and Data Analysis}
\label{sec:obs_data}

To model the new possible source, we add a source at the location of the highest probability pixel\footnote{The skymap was obtained from \url{https://dcc.ligo.org/LIGO-P1800381-v6/public/GW170104_skymap.fits.gz} (accessed on 01/13/2025).}, R.A = 47.283$^{\circ}$ and Dec = -44.59$^{\circ}$, corresponding to the maximum-probability localization of the GW170814 event as estimated by the LIGO and Virgo Collaborations.
This source is modeled using a spectral FileFunction\footnote{\url{https://fermi.gsfc.nasa.gov/ssc/data/analysis/scitools/source_models.html\#FileFunction} (accessed on 01/13/2025).}, implemented in \texttt{FERMIPY}, with the normalization parameter kept fixed. The spectrum is computed using the open-source public code \texttt{BlackHawk} \cite{2019EPJC79_693A} and we normalized the signal for a total mass emitted in morsels of 1 $M_\odot$, assuming that all morsels have the same mass, and for a merger at a distance D = 540 Mpc from Earth (corresponding to GW170814). 
We consider a range of morsel masses, $m_{\text{BH}} = 1.8 \times 10^{8} \text{kg}, 2.5 \times 10^{8} \text{kg}, 3.0 \times 10^{8} \text{kg}, 3.5 \times 10^{8} \text{kg}, \text{and}~4.0 \times 10^{8} \text{kg}$. These morsels have a corresponding evaporation time of $t_{\text{evap}} = 29, 79, 139, 225, \text{and}~339$ days respectively. These particular values were not chosen based on specific assumptions related to the GW170814 event, but rather correspond to the set of spectra available from the simulations employed.

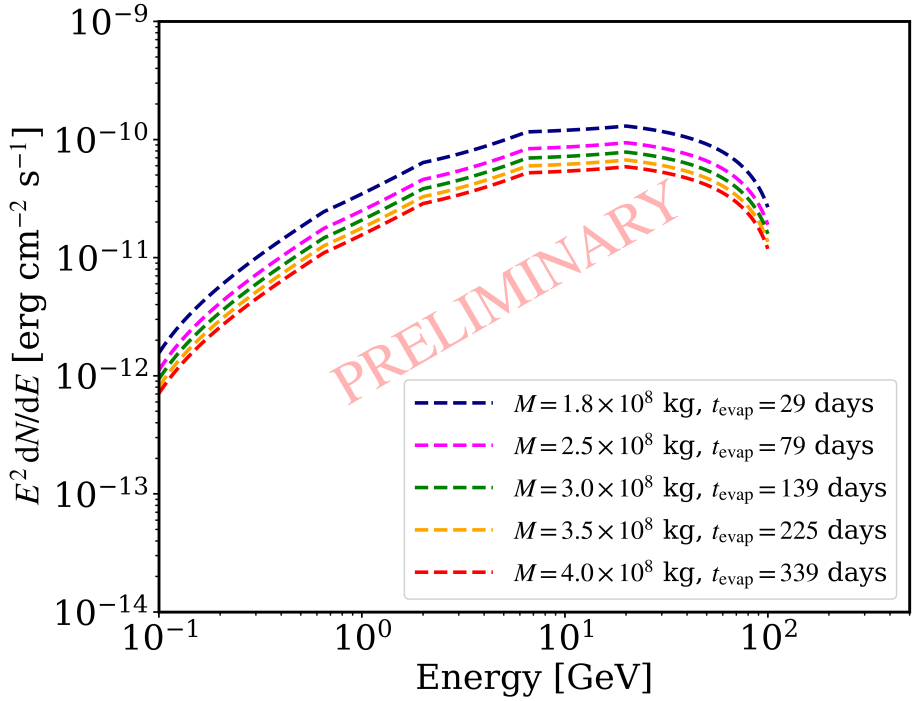
\begin{figure}[t]
\centering
\includegraphics[width=0.8 \linewidth]{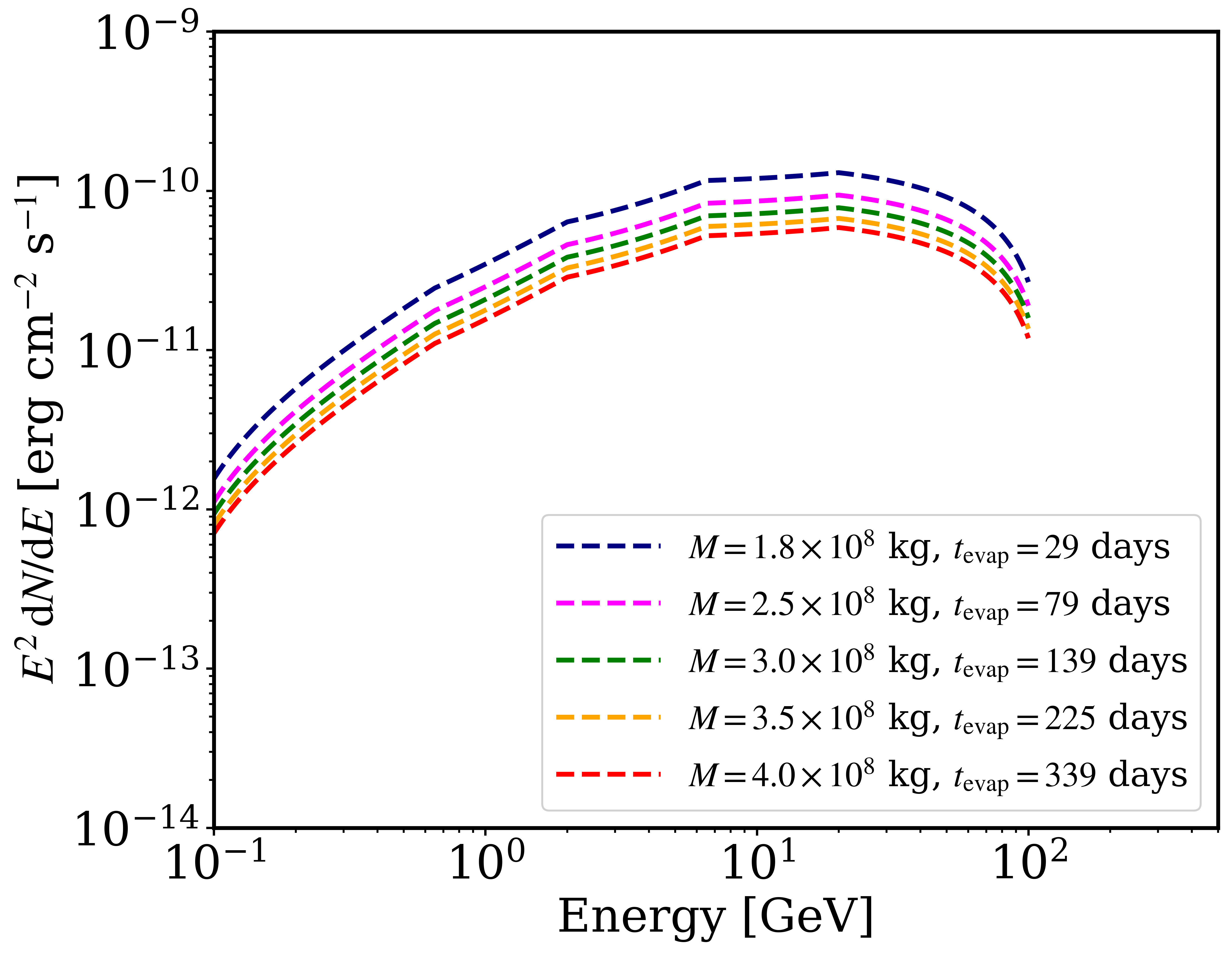}
\begin{tikzpicture}[overlay]
    \node[rotate=30, text opacity=0.3, scale=2, text=red] at (-5.5,5.5) {PRELIMINARY};
\end{tikzpicture}
\caption{The simulated differential flux spectrum from one solar mass of \emph{black hole morsels} of individual masses , $m_{\text{BH}} = 1.8 \times 10^{8} \text{kg}, 2.5 \times 10^{8} \text{kg}, 3.0 \times 10^{8} \text{kg}, 3.5 \times 10^{8} \text{kg}, \text{and}~4.0 \times 10^{8} \text{kg}$, at a distance of 540 Mpc from Earth. These spectra are averaged over the entire morsel lifetime. These spectra are used to model the possible new source at the location of the highest probability pixel of GW170814.}
\label{fig:input_spectra}
\end{figure}

We investigate the month-by-month spectra and for each month until $t_{\text{evap}}$, we consider the spectral values obtained from the corresponding simulations. We also consider only the until the first month after $t_{\text{evap}}$, and here attenuate the spectra by $\Delta t$/30, where $\Delta t = t_{\text{obs}} - t_{\text{evap}}$. More specifically, this means that if a (hypothetical) GW event were to happen on Jan 1st and the \emph{black hole morsels} have an evaporation time of 79 days, we use the simulated spectra for January and then separately for February and then for March we would take that spectrum multiplied (19/30). The input spectra considered for each of the five morsel masses, for $t_{\text{obs}} \leq t_{\text{evap}}$ (i.e until evaporation), is shown in Fig.~\ref{fig:input_spectra}.

The contributions from the isotropic and Galactic diffuse backgrounds were modeled using the most recent templates for isotropic and Galactic diffuse emission, iso\_P8R3\_SOURCE\_V2\_v1.txt and gll\_iem\_v07.fits respectively.
Sources in the 4FGL-DR3 catalog \citep{4fgl_dr3} within a radius of $20^{\circ}$ from the location of the most probable pixel were included in the model with their spectral parameters fixed to their catalog values. This takes into account the gamma-ray emission from sources lying outside the RoI which might yet contribute photons to the data, especially at low energies, due to the size of the point spread function of the \textit{Fermi}-LAT. The normalization factors for both diffuse templates, together with the spectral normalization of all modeled sources within the RoI, were left free. Furthermore, the spectral shape parameters of all modeled sources within $3^{\circ}$ of the most probable pixel were left free to vary, while those of the remaining sources were fixed to the values reported in the 4FGL-DR3 catalog. The \textit{gtfindsrc} routine was also applied to search for any additional point sources present in the model and not included in the 4FGL-DR3 catalog. A binned maximum likelihood analysis was then performed for all time intervals i.e for each of the considered morsel masses and with each month considered individually, using a spatial binning of 0.1$^{\circ}$ pixel$^{-1}$ and two energy bins per decade.

\section{Discussion}
\label{sec:discuss}

The significance of the gamma-ray emission for each time interval was evaluated using the maximum likelihood test statistic (TS). The TS is defined as the log-likelihood ratio between the maximized likelihoods with and without an additional source at the specified location, $L_{1}$ and $L_{0}$ respectively \cite{RN7}:
\begin{equation}
   \text{TS} = -2 \ln \left( \frac{L_{0}}{L_{1}} \right) \, .
	\label{eq:1}
\end{equation}

For a fixed morsel mass, the additional source has no free spectral parameters, as the spectrum is fully determined by the expected Hawking radiation. We find TS values $\leq 0$ for all tested time bins and black hole morsel masses, which clearly indicates that the inclusion of this additional source decreases the overall likelihood of the ROI model. In the following, we quantify this null result and translate it into exclusion limits on the parameter space of black hole morsels.

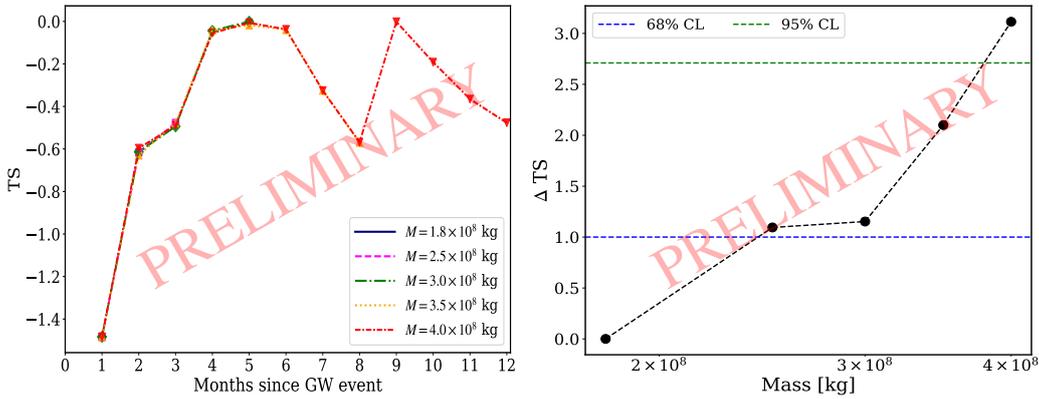
\begin{figure*}[h]
\centering
\includegraphics[width=0.45\linewidth,height=0.35\linewidth]{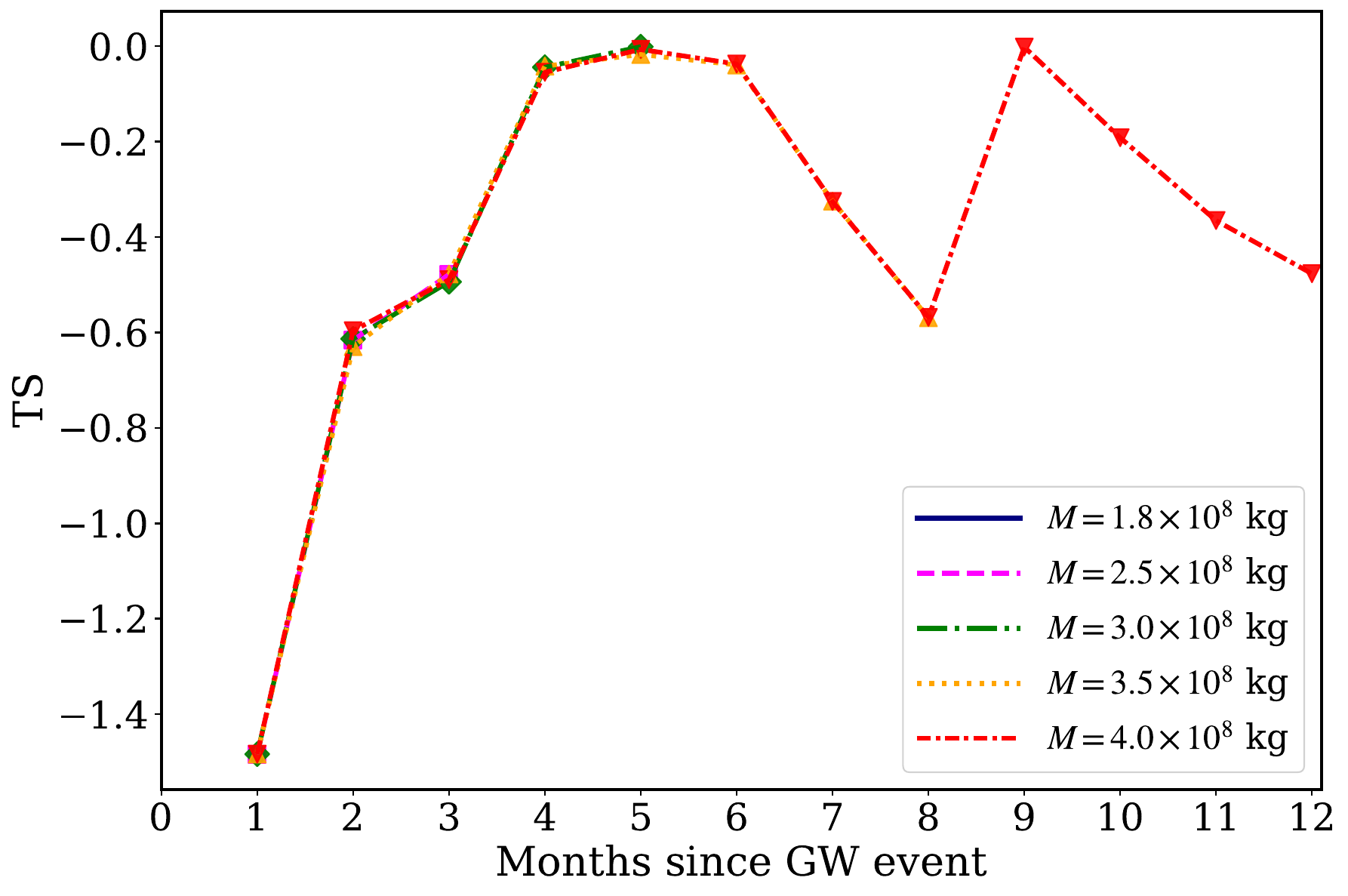}
\includegraphics[width=0.45\linewidth,height=0.35\linewidth]{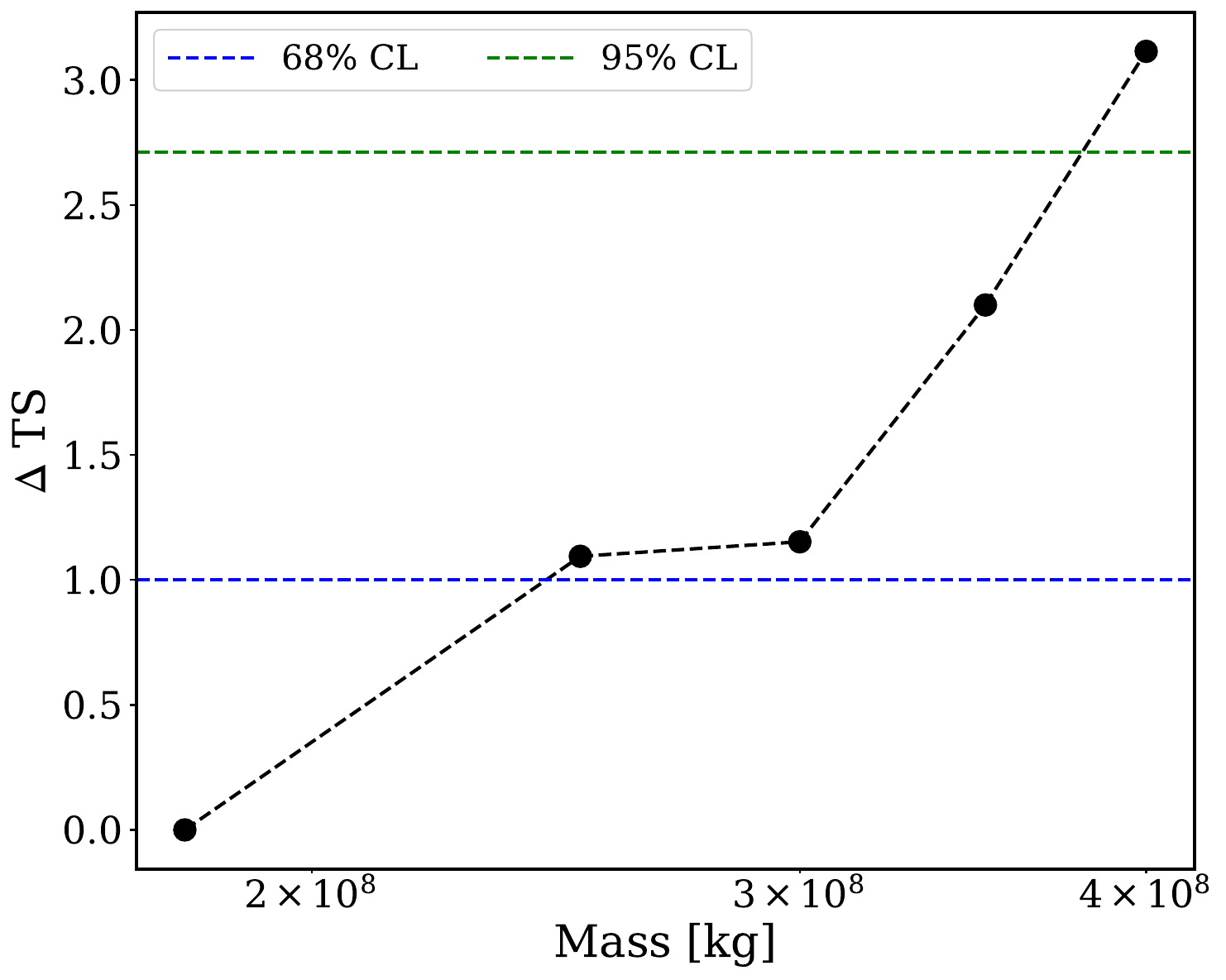}

\begin{tikzpicture}[overlay]
    \node[rotate=30, text opacity=0.3, scale=2, text=red] at (-2.75,3.5) {PRELIMINARY};
    \node[rotate=30, text opacity=0.3, scale=2, text=red] at (4,3.5) {PRELIMINARY};
\end{tikzpicture}
\caption{
\textbf{Left:} Monthly test statistic (TS) values from \textit{Fermi}-LAT data at the GW event location for black hole morsel models of different masses. All curves show negative TS across the 12-month window, indicating no significant gamma-ray counterpart. 
\textbf{Right:} $\Delta \mathrm{TS}$ as a function of black hole mass, computed relative to the least negative TS case ($1.8 \times 10^{8}\,\mathrm{kg}$). The horizontal dashed lines indicate exclusion thresholds at 68.3\% ($\Delta \mathrm{TS} = 1$) and 95\% ($\Delta \mathrm{TS} = 2.71$) confidence levels. The $4 \times 10^{8}\,\mathrm{kg}$ scenario exceeds the 95\% threshold and is therefore excluded at this level.
}
\label{fig:ts_month}
\end{figure*}

The TS values obtained over each month for the considered \emph{black hole morsel} masses are shown in Fig.~\ref{fig:ts_month} (left panel). All curves show negative TS across the 12-month window, indicating no significant gamma-ray counterpart. The curves terminate at the corresponding evaporation time $t_{\text{evap}}$ of each mass scenario (for example $29$ days for $1.8 \times 10^{8}$\,kg, $139$ days for $3.0 \times 10^{8}$\,kg). Beyond $t_{\text{evap}}$, no signal is expected. The curves nearly overlap because the additional source has no free spectral parameters beyond the fixed spectral template, so the resulting TS values are almost identical across the tested masses.

While our study does not provide support for the existence of \emph{black hole morsels}, it is nevertheless possible to use the obtained upper limits to provide exclusion limits on their masses. 
Taking the least negative TS value as a reference and computing $\Delta$TS for other cases shows that the $4 \times 10^{8}$\,kg scenario can be excluded at the 95\% confidence level, as seen in Fig.~\ref{fig:ts_month} (right panel). Here the confidence levels are derived from the standard likelihood ratio test, where $\Delta \mathrm{TS} = 1$ and $\Delta \mathrm{TS} = 2.71$ correspond to the 68.3\% and 95\% exclusion thresholds, respectively. Further exclusion limits will require extending the mass grid on both sides and incorporating the late-time gamma-ray spike expected near evaporation, a task that is the subject of ongoing work and will be presented in a forthcoming publication.

Looking ahead, the methodology developed here can be extended in several promising directions. First, the late-time gamma-ray spike expected near the final stages of morsel evaporation will be incorporated into our templates, potentially improving sensitivity. Second, future high-energy facilities such as the Cherenkov Telescope Array Observatory (CTAO; \cite{2019scta.book.....C}) will significantly expand the accessible parameter space and may enable the detection of weaker signals. In addition, one could test a broader mass range and a distribution of produced morsel masses, which would likely provide a more realistic description.
Finally, applying this search strategy systematically across the growing catalog of binary black hole mergers observed by LIGO--Virgo--KAGRA will provide increasingly stringent constraints, and perhaps eventually the first observational confirmation of Hawking radiation.

\section*{Acknowledgements}
The \textit{Fermi}-LAT Collaboration acknowledges generous ongoing support from a number of agencies and institutes that have supported both the development and the operation of the LAT as well as scientific data analysis. These include the National Aeronautics and Space Administration and the Department of Energy in the United States, the Commissariat à l'Energie Atomique and the Centre National de la Recherche Scientifique / Institut National de Physique Nucléaire et de Physique des Particules in France, the Agenzia Spaziale Italiana and the Istituto Nazionale di Fisica Nucleare in Italy, the Ministry of Education, Culture, Sports, Science and Technology (MEXT), High Energy Accelerator Research Organization (KEK) and Japan Aerospace Exploration Agency (JAXA) in Japan, and the K. A. Wallenberg Foundation, the Swedish Research Council and the Swedish National Space Board in Sweden.

Additional support for science analysis during the operations phase is gratefully acknowledged from the Istituto Nazionale di Astrofisica in Italy and the Centre National d'Etudes Spatiales in France. This work performed in part under DOE Contract DE- AC02-76SF00515.

\end{document}